# Two-Dimensional Band Dispersion of Ultra-Flat Hexagonal Bismuthene Grown on Ag(111) Bulk and Quantum-Well Films

Kazutoshi Takahashi *, Masaki Imamura, Haruto Ikeda, Ryosuke Koyama, Isamu Yamamoto, and Junpei Azuma
*Synchrotron Light Application Center, Saga University, Saga 840-8502, Japan*
(Dated: September 27, 2023)

Two-dimensional band dispersion of (2×2) superstructure with Bi grown on Ag(111), which has been proposed as an ultraflat hexagonal bismuthene, is investigated using angle-resolved photoemission spectroscopy (ARPES). The (2×2)-Bi superstructure can be grown on the Ag(111) surface at low temperatures; it transforms into a surface alloy with a ($\sqrt{3}\times\sqrt{3}$) superstructure at 300 K. ARPES measurements are in agreement with the band structure of ultraflat bismuthene in previous reports. The band structure of (2×2)-Bi surface remains unchanged with decreasing Ag layer thickness, indicating the limited penetration of Bi p-orbitals into the Ag layer.

## I. INTRODUCTION

Two-dimensional topological insulators (2D TIs) with large bulk gaps exhibit one-dimensional topological edge states that can host dissipationless transport and are potential candidates for realizing quantum spintronic devices [1,2]. Group-V elemental 2D materials are attracting increasing attention because of their intriguing properties, such as tunable band gap, high carrier mobility, and promising transition to a topological non-trivial phase. As a heavy element, Bi induces stronger spin-orbit coupling and a larger bandgap opening of the TI state. Monolayer Bi mainly exhibits two structural forms: α-phase and β-phase. The α-bismuthene (α-Bi) has a black phosphorus-like A17 structure that crystallizes in the (110) orientation with a rectangular unit cell and a paired bilayer (BL) structure. The 1- and 2-BL α-Bi transform into non-trivial 2D TIs in the absence of buckling [3]. Dirac-like states are induced around $\bar{X}_1$ and $\bar{X}_2$ points of α-Bi and α-antimonene by glided mirror symmetry and SOC [4–6]. Hexagonal (111) β-bismuthene (β-Bi) is also classified as a paired BL owing to the presence of atoms at the upper and lower sites of the buckled hexagonal layer. β-Bi has been reported to be a 2D TI with a well-localized topological edge state [7,8]; this is experimentally confirmed despite the complexity due to hybridization between the surface states of $Bi_2Te_3$ substrate and the Bi(111). Interaction with the substrate is vital for tuning the material properties through symmetry, electronic coupling, and strain. A monolayer of Bi atoms arranged in a planar honeycomb geometry on SiC(0001) is a novel platform to achieve the topological edge channel with a large bulk gap [9,10]. Scanning tunneling spectroscopy (STS) measurements of planar Bi grown on SiC(0001) have demonstrated a metallic edge state and a large band gap in the bulk area. The large band gap of flat Bi on SiC originates from the $p_x$ and $p_y$ orbitals after the π-bond contribution, whereas the $p_z$ orbital is pushed out from near the Fermi level owing to hybridization with the substrate [9].

Recently, another flat honeycomb structure of bismuthene was distinguished on Ag(111), which was grown and kept at a low temperature [11]. The flat bismuthene on Ag(111) exhibit a (2×2) superstructure with Bi atoms located at the hollow sites of Ag(111), indicating that the distance between two Bi atoms on Ag(111) is 3.34 Å, which is 108.7% of the interatomic distance in the buckled Bi(111) layer (3.07 Å). The band calculation showed that a large SOC-induced gap was opened at the $\bar{K}$ point, as the Bi $p_z$ orbital was moved away from the Fermi level owing to selective coupling with the Ag s orbital [11]. Although the STS measurement revealed the metallic density of states (DOS) at zig-zag and armchair edges near the Fermi level, knowledge of the electronic band structure in a wide energy and momentum range is still necessary to understand the intriguing electronic states of the ultraflat-Bi structure.

In this study, we examined the two-dimensional band dispersion of (2×2)-Bi superstructure grown on Ag(111) bulk and quantum-well (QW) films via angle-resolved photoemission spectroscopy (ARPES) using synchrotron radiation. The band structure of (2×2)-Bi on Ag(111) showed excellent agreement with the band calculation of flat bismuthene, where Bi $p_{xy}$ orbitals resulted in a large topological gap of approximately 1.4 eV at the $\bar{K}$ point. The band structure of (2×2)-Bi did not change when the thickness of the Ag layers was decreased to 4 monolayer (ML), indicating that the penetration of Bi p-orbitals into the Ag layer was small. The asymmetry and chemical shift of the Bi 5d core-level peaks of the (2×2)-Bi surface are smaller than those of the ($\sqrt{3}\times\sqrt{3}$)-Bi surface where the Bi atoms are embedded in the top Ag layer.

## II. EXPERIMENT

All measurements were performed at the plane grating monochromator (PGM) station of the Saga University Beamline (beamline 13) at the Saga Light Source [12]. Highly B-doped Si(111) substrates were flashed at 1470 K for 5 s several times and kept at 1270 K for 5 min. Si(111) (√3×√3)-B surfaces were prepared by slow cooling to approximately 300 K. Ag films with thicknesses of 110, 11, and 4 ML were prepared by depositing Ag onto the Si substrate at 100 K. The deposition rate was approximately 0.5 ML/min. The sample was then annealed at 300 K to yield a well-ordered Ag film. The deposition of Bi was performed at 100 K with a deposition rate of 0.05 ML/min. In this study, the deposition of 1-ML is defined as the density of Ag atoms in an Ag(111) plane (13.8 atoms nm$^{-2}$). The deposition rates of Ag and Bi were calibrated using the binding energies of the QW states in Ag(111)/Si(111) (7×7) and Bi(111)/Si(111) (7×7) ultrathin films, respectively [13,14]. ARPES was performed using an MB Scientific A-1/Lens4 analyzer with horizontally polarized photons at 19 eV with a spot size of 0.5 mm in diameter at 40 K. The typical photon flux on the sample was approximately $6×10^{10}$ photons/s. The Fermi energy and energy resolution were confirmed using the measurements for the gold reference, and the overall energy resolution was estimated to be 43 meV.

## III. RESULTS AND DISCUSSION

Figure 1(a) shows the low-energy electron diffraction (LEED) pattern for the Bi deposition of 0.16 ML on Ag(111) films of 11 ML at 100 K. In addition to the hexagonal pattern corresponding to the Ag(111) surface, new hexagonal diffraction spots were clearly observed; they show a (2×2) superstructure with respect to the Ag(111) surface lattice, consistent with the STM result of the previous study [11]. LEED measurements were also performed after the photoemission measurement without elevating the sample temperature. The unchanged (2×2) spots were clearly observed after the photoemission measurement. Figure 1(b) shows the LEED pattern after annealing at 300 K, where the formation of a Ag$_2$Bi alloy with a (√3×√3) superstructure [15] is indicated. The Bi amount of 0.16 BL is insufficient for the 0.50 ML corresponding to the structural model of the flat bismuthene with every Bi atoms located at the hollow sites of Ag(111). However, further deposition of Bi induces an extra LEED spot corresponding to the Bi(110) layer, which has been reported by Zhang et al. [16]. The additional Bi(110) layer probably grew on the (2×2)-Bi structure before completing (2×2) region in whole area of the Ag(111) surface.

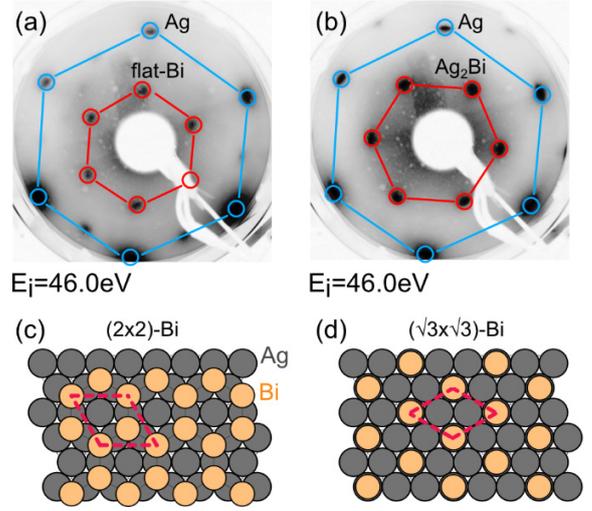

FIG. 1. LEED patterns of the Bi deposition of 0.16 ML on Ag(111) films of 11 ML at 100 K (a) and those after annealing at 300 K (b). The atomic arrangement of (2×2) (c) and (√3×√3) (d) superstructures with respect to the Ag(111) surface lattice.

Figure 2(a) shows the surface Brillouin zones (SBZs) of the (2×2)-Bi structure and Ag(111) surface. ARPES along the $\bar{\Gamma}$-$\bar{K}$ line of the 110-ML-thick Ag(111) film on the Si(111) (√3×√3)-B surface is shown in Fig. 2(b). The intense feature immediately below the Fermi level is the Shockley surface state. In addition to the broad background intensity due to the indirect transition from the sp-band, a structure of the sp-band direct transition is observed; it crosses the Fermi level at the wave number of 1.1 Å$^{-1}$ when a photon energy of 19 eV is used. As shown in Fig. 2(c), the Shockley surface state disappears, and the intensity near the Fermi level at the $\bar{\Gamma}$ point also decreases after the growth of the (2×2)-Bi surface on the 110-ML-thick Ag(111) film. The structure attributed to the sp-band direct transition is observed at the same position, whereas its intensity is slightly decreased. Notably, a band dispersing to the high binding energy side with tops at the first and second $\bar{K}$ points appears on the (2×2)-Bi surface. The observed band significantly differs from the well-established Rashba split bands on the Ag$_2$Bi alloy with the (√3×√3) superstructure [15,17].

The observed dispersion can be compared with the band structures of ultraflat bismuthene from the theoretical calculation by Sun et al. [11]. The calculation of freestanding bismuthene with SOC shows that an SOC-induced gap appears near the Fermi level, with the upper and lower bands consisting of p$_{xy}$ and p$_z$ orbitals, respectively. The freestanding bismuthene shows p$_z$-derived bands located immediately below and above the Fermi level. When the underlying Ag layer with a thickness of 1 ML is included to cause substrate-orbital-filtering effect in the calculation, the strong coupling with the Ag s orbital pushes the Bi $p_z$ orbital away from the Fermi level to the conduction bands. The remaining $p_{xy}$ bands form a large band gap of 1 eV at the $\bar{K}$ point, which increases to approximately 1.4 eV when the 3 ML Ag(111) layers are considered. The lower band is located at approximately 0.5 and 1.0 eV below the Fermi level at $\bar{K}$ point when the 1 and 3 ML Ag(111) layers are included, respectively. The system with the underlying Ag(111) layers is calculated to be topologically non-trivial. Our ARPES shows that the (2×2)-Bi-induced band is located at 1.24 eV at both the first and second $\bar{K}$ points, as shown in Fig. 2(c). The band disperses to a higher binding energy from $\bar{K}$ point to $\bar{M}$ point. The observed dispersion is consistent with the calculated band with $p_{xy}$ character in ultraflat-Bi of the (2×2) structure on 3 ML Ag(111) layers.

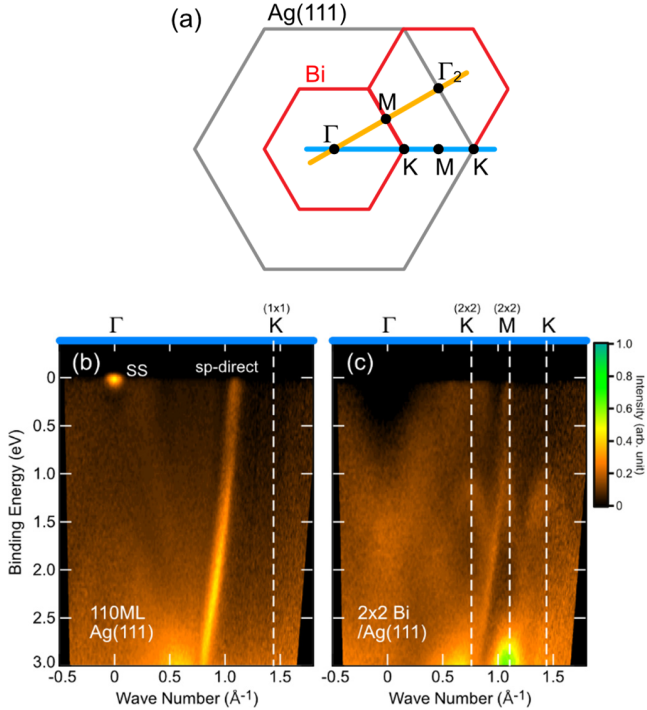

FIG. 2 Surface Brillouin Zones (SBZs) of Ag(111) surface and (2×2)-Bi superstructure. (b) Angle-resolved photoemission spectroscopy (ARPES) intensity map along $\bar{\Gamma}$-$\bar{K}$ line of 110ML-thick Ag(111) film. (c) ARPES intensity map along $\bar{\Gamma}$-$\bar{K}$-$\bar{M}$-$\bar{K}$ line of (2×2)-Bi surface.

This study attempted to probe the penetration depth of the Bi sp states into Ag layers using measurements on (2×2)-Bi structures grown on Ag(111) QW films. Figures 3(a) and (b) show the ARPES along $\bar{\Gamma}$-$\bar{K}$ and $\bar{\Gamma}$-$\bar{M}$ lines of the 11 ML Ag(111) film, respectively. In addition to the Shockley state observed immediately below the Fermi level at $\bar{\Gamma}$ point, quantized Ag sp-bands with parabolic dispersions are clearly observed, indicating the excellent crystallinity and thickness homogeneity of the Ag(111) film. Figures 3(c) and (d) show the ARPES along $\bar{\Gamma}$-$\bar{K}$-$\bar{M}$ and $\bar{\Gamma}$-$\bar{M}$-$\bar{\Gamma}$ lines of the (2×2)-Bi structure on Ag(111) QW films with a thickness of 11 ML. On the (2×2)-Bi structure, SS disappears and QW states shift to lower binding energies. The (2×2)-Bi induced band is located at 1.64 eV at the second $\bar{\Gamma}$ point, as shown in Fig. 3(d), which is consistent with the energy position calculated by Sun et al. [11]. Figures 3(e) and (f) show the ARPES of the (2×2)-Bi structure on the 4-ML-Ag(111) QW film. The Bi $p_{xy}$ band at the $\bar{K}$ and $\bar{\Gamma}$ points is located at the same binding energy as that of the 11 ML thick Ag(111). The unchanged energy positions of the occupied Bi $p_{xy}$ states indicate the small penetration depth of Bi states into the Ag film. The penetration depth is lower than 4 ML.

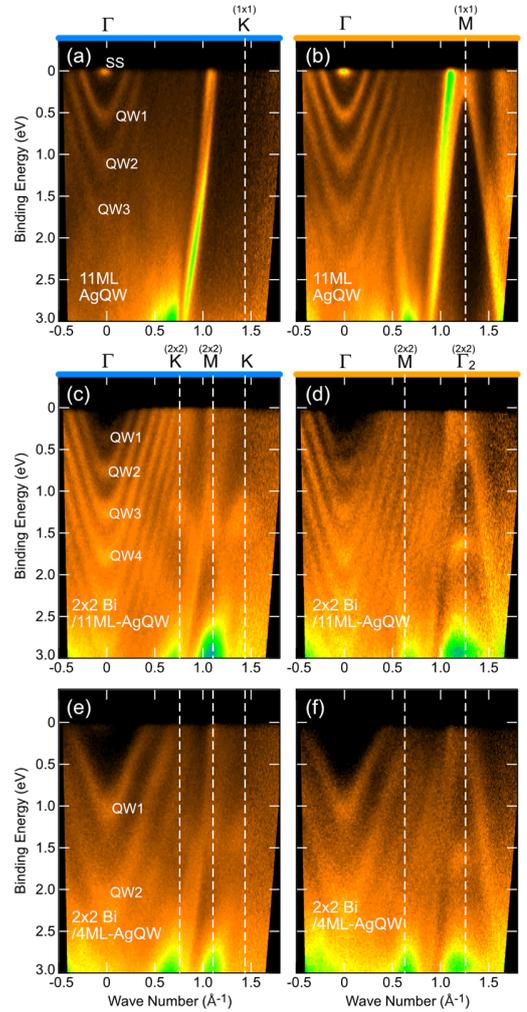

FIG.3. ARPES intensity map along the $\bar{\Gamma}$-$\bar{K}$ (a) and $\bar{\Gamma}$-$\bar{M}$ (b) lines of 11-ML-thick Ag(111) QW film. ARPES intensity map along the $\bar{\Gamma}$-$\bar{K}$-$\bar{M}$ (c) and $\bar{\Gamma}$-$\bar{M}$-$\bar{\Gamma}$ (d) lines of the (2×2)-Bi structure on 11-ML-thick Ag(111) QW film. ARPES intensity map along the $\bar{\Gamma}$-$\bar{K}$-$\bar{M}$ (e) and $\bar{\Gamma}$-$\bar{M}$-$\bar{\Gamma}$ (f) lines of the (2×2)-Bi structure on 4-ML-thick Ag(111) QW film.

Bi 5d and Ag 4d spectra of the (2×2)-Bi surface grown on 11- and 4-ML-thick Ag films are shown in Fig. 4(a). The spectra before Bi deposition are also shown in Fig. 4(a). Ag 4d spectra on 11- and 4-ML-thick films exhibit the same shape. The Ag 4d spectra show a decrease in intensity and a small change in shape upon the deposition of Bi; however, their peak positions do not change. Details of the Bi $5d_{5/2}$ peaks are shown in Fig. 4(b). The Bi 5d peaks of (2×2)-Bi on Ag(111) films show apparent asymmetry, displaying a high binding energy tail. The core-level spectrum gives a direct signature of the metallic or semiconducting/insulating nature of the sample under study through a characteristic asymmetric line shape for metallic bands [18]. The asymmetric tails toward high binding energies reflect gapless excitations for metallic systems. The asymmetric nature of the Bi core-level spectra indicates that Bi contributes to the density of states around the Fermi energy, and the Doniach–Šunjić functions are used in the line shape analysis. Using six parameters, peak height, binding energy, branching ratio (b.r.), Gaussian width (GW), Lorentzian width (LW), and singularity index α, the Bi 5d spectra are well-fitted by one 5d component after subtraction of an integration-type background.

TABLE 1. Fitting parameters for Bi 5d core-level spectra using the Doniach–Šunjić functions.

|  | BE | b.r. | GW | LW | α |
|---|---|---|---|---|---|
| Bi(111) | 23.886 | 0.668 | 0.333 | 0.157 | 0.026 |
| 2×2 | 23.746 | 0.678 | 0.226 | 0.211 | 0.057 |
| √3×√3 | 23.630 | 0.739 | 0.168 | 0.212 | 0.090 |

The peak position and asymmetry index in the line-shape analysis exhibit large differences among the three surfaces. The singularity index was fitted with an accuracy of ±0.004. A thick Bi(111) film on a Si(111) substrate is a semimetal with only a small overlap of the valence and conduction bands. The asymmetry index of the Bi(111) film is as small as 0.026, indicating only a small contribution of Bi to the density of states at the Fermi level. The asymmetry index at the (2×2)-Bi surface is 0.057, reflecting the interaction between the Ag sp and Bi sp states; however, it is smaller than the asymmetry index at the (√3×√3) surface, where one-third of the Ag atoms in the Ag(111) surface are substituted by Bi atoms to form a $Ag_2Bi$ surface alloy. As Bi has a greater electronegativity than Ag, electron transfer from Ag to Bi is expected during alloy formation. The electron transfer to Bi shifts the Bi core level to the lower binding energy side. The binding energy of the Bi 5d level at the (2×2)-Bi surface is on the lower binding energy side than that at Bi(111), indicating electron transfer from Ag to Bi. However, the shift is smaller than that at the (√3×√3) structure forming the $Ag_2Bi$ alloy. The smaller asymmetry and chemical shift of the Bi core levels on the (2×2) surface than those on the (√3×√3) surface indicate a smaller contribution of Bi states at the Fermi energy. The smaller electronic coupling on the (2×2) surface is also consistent with the reported structure [11], where Bi atoms are arranged at the hollow sites of Ag(111) with (2×2) periodicity without substituting Ag atoms.

The epitaxial growth of antimonene, another Group-V 2D material, has also been explored using Ag(111) and Cu(111) substrates. Buckled monolayers with larger lattice constants than the freestanding form were identified on alloyed first layers, which were grown at room temperature followed by mild annealing [14,15]. Previous experiments concluded that both the first and second layers of Sb were grown as bulked hexagonal structures without alloying [21,22]. Recently, Zhang *et al*. argued that a large-area-occupying and atomically flat $Cu_2Sb$ and $Ag_2Sb$ alloy could form on the surfaces after annealing at an appropriate temperature, thus supporting the growth of pseudo-cubic antimonene with significant strain [23]. These experiments were not performed at decreased substrate temperatures and

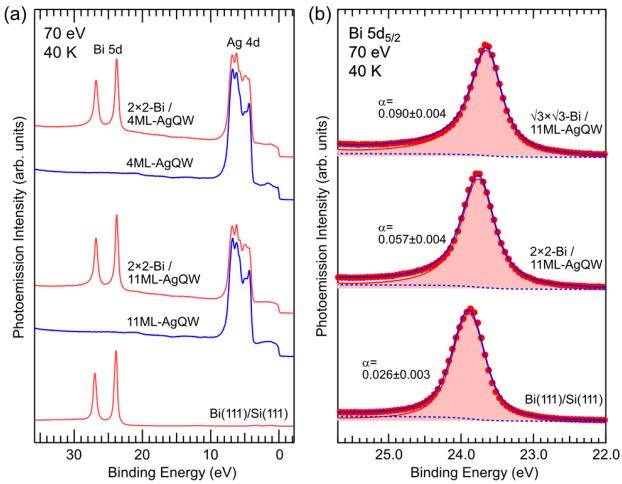

FIG.4 (a) Core-level photoemission spectra of thick Bi(111) film grown on Si(111)-7x7 surface, 11ML and 4ML-thick Ag(111) quantum-well films, and (2×2)-Bi structure on Ag quantum-well films measured using a photon energy of 70 eV. (b) Line shape analysis of Bi $5d_{5/2}$ core-level spectra using the Doniach–Šunjić functions.

exhibited ($\sqrt{3} \times \sqrt{3}$) periodicity. On the other hand, (2×2)-Bi on Ag(111) was grown and kept at a low temperature. The observed band dispersions and small asymmetry of the core-level peak strongly support a non-alloyed (2×2)-Bi layer with extensive tensile strain and a large topological band inversion enabled by the substrate-orbital-filtering effect proposed by Sun *et al.* [11].

## IV. CONCLUSION

The electronic structure of ultraflat hexagonal bismuthene with a (2×2) superstructure grown on an Ag(111) surface was elucidated via photoemission spectroscopy using synchrotron radiation. The asymmetry of the Bi core level of (2×2) surface was smaller than that of the ($\sqrt{3} \times \sqrt{3}$) surface alloy, indicating that the interaction between the Ag sp and Bi sp states on (2×2)-Bi surface is small. The band structure of (2×2)-Bi on Ag(111) did not change when the thickness of the Ag layers was decreased to 4 ML, indicating that the penetration of Bi p-orbitals into the Ag layer was small. The band structure of (2×2)-Bi showed excellent agreement with the band calculation of ultraflat bismuthene, where Bi $p_{xy}$ orbitals resulted in a large topological gap of approximately 1.4 eV at the $\bar{K}$ point.

## ACKNOWLEDGMENTS

This work was supported by JSPS KAKENHI (Grant No.: 20K03821) and the Partnership Project for Fundamental Technology Research of the Ministry of Education, Culture, Sports, Science and Technology of Japan.

* ktaka@cc.saga-u.ac.jp